\documentclass[sigconf]{acmart}
\usepackage{hyperref}

\begin{document}

%
\title{Predicting Patent Citations to measure Economic Impact of Scholarly Research}

%
\author{Abdul Rahman Shaikh}
\email{z1841128@students.niu.edu}

\affiliation{%
  \institution{Northern Illinois University}
  \streetaddress{1425 W. Lincoln Hwy.}
  \city{DeKalb}
  \state{IL}
  \postcode{60115-2828}
}

\author{Hamed Alhoori}
\email{alhoori@niu.edu}
\affiliation{%
  \institution{Northern Illinois University}
  \streetaddress{1425 W. Lincoln Hwy.}
  \city{DeKalb}
  \state{IL}
  \postcode{60115-2828}
}
%
\renewcommand{\shortauthors}{Shaikh and Alhoori}

%
\begin{abstract}
A crucial goal of funding research and development has always been to advance economic development. On this basis, a considerable body of research undertaken with the purpose of determining what exactly constitutes economic impact and how to accurately measure that impact has been published. Numerous indicators have been used to measure economic impact, although no single indicator has been widely adapted.  Based on patent data collected from Altmetrics we predict patent citations through various social media features using several classification models. Patents citing a research paper implies the potential it has for direct application in its field. These predictions can be utilized by researchers in determining the practical applications for their work when applying for patents.
\end{abstract}

%
\keywords{Patents, Economic Impact, Social Media, Altmetrics}

%
\maketitle

\section{Introduction}
The main purpose of the patent system is to stimulate innovation in the market. Given that relevant data are readily available, patents and patent statistics are widely used by researchers to identify and explore areas of technical change and innovation simultaneously analyzing economic growth. Lv et al. \cite{Lv} identified 161 converging technologies by performing cluster analysis on USPC class patents of five parties for 10 years from 2005 to 2015. Langinier and Moschini \cite{Langinier} found that protecting innovations through patents is a crucial task for technical improvements in the industry and patents produce innovations that stimulate economic growth. Patent citations have great potential in terms of providing a way to measure economic impact through indicators of patent quality. Squicciarini et al. \cite{Squicciarini} provided various indicators to measure patent quality. Specifically, they found that indicators such as patent family size, patent citations, patent renewals, and claims provide information pertinent to the technological and economic value of innovations. Further, they proposed a patent quality index based on four to six dimensions of patent quality. Based on an analysis of patents and their citations for the period of 1963 to 1995, Hall et al. \cite{Hall} found that patent citations include valuable information regarding the market value of firms, R\&D and patent counts. 

Patent data and patent citations are very useful indicators of the inventions and the R\&D expenditures of a firm. Based on a survey of previous studies on patent statistics, Griliches \cite{Griliches} found a very strong relationship between patents and a firms R\&D expenditures in cross-sectional dimensions. Ribeiro et al. \cite{Ribeiro:2014:MUG:2670793.2670802} examined 167,315 United States Patent and Trademark Office (USPTO) patents from 2009 and the papers cited in those patents and found that a global knowledge flow exists between universities and firms Van Raan \cite{vanRaan17} studied the background of scientific non-patent references (SNPR) and concluded that patents with a high economic value invention were cited highly. Patent quality can be assessed by multiple indicators of economic value such as patent claims, SNPRs, patent family size, patent renewal, and forward citations of patents. 

Predicting amount of patent citations can be helpful in measuring the economic impact of research and in understanding how knowledge is commercialized. Our study built classifiers to predict the likelihood of a research article being cited in patents using social media features.

\section{Data Collection}
A random dataset was collected from altmetric.com which initially contained a million records. We only considered the articles published after 2010 since those records would have higher social media mentions. We performed data cleaning on the dataset to look for duplicate records and null values ending up with 784,665 data records. From these records 372,755 records were cited by patents and 411,910 records were not cited by patents. We removed few social media features such as Weibo, F1000, Q\&A and Reddit since they mostly had null values for most records.

\begin{figure}
    \centering
    \vspace{1em}
    \includegraphics[width=0.5\textwidth]{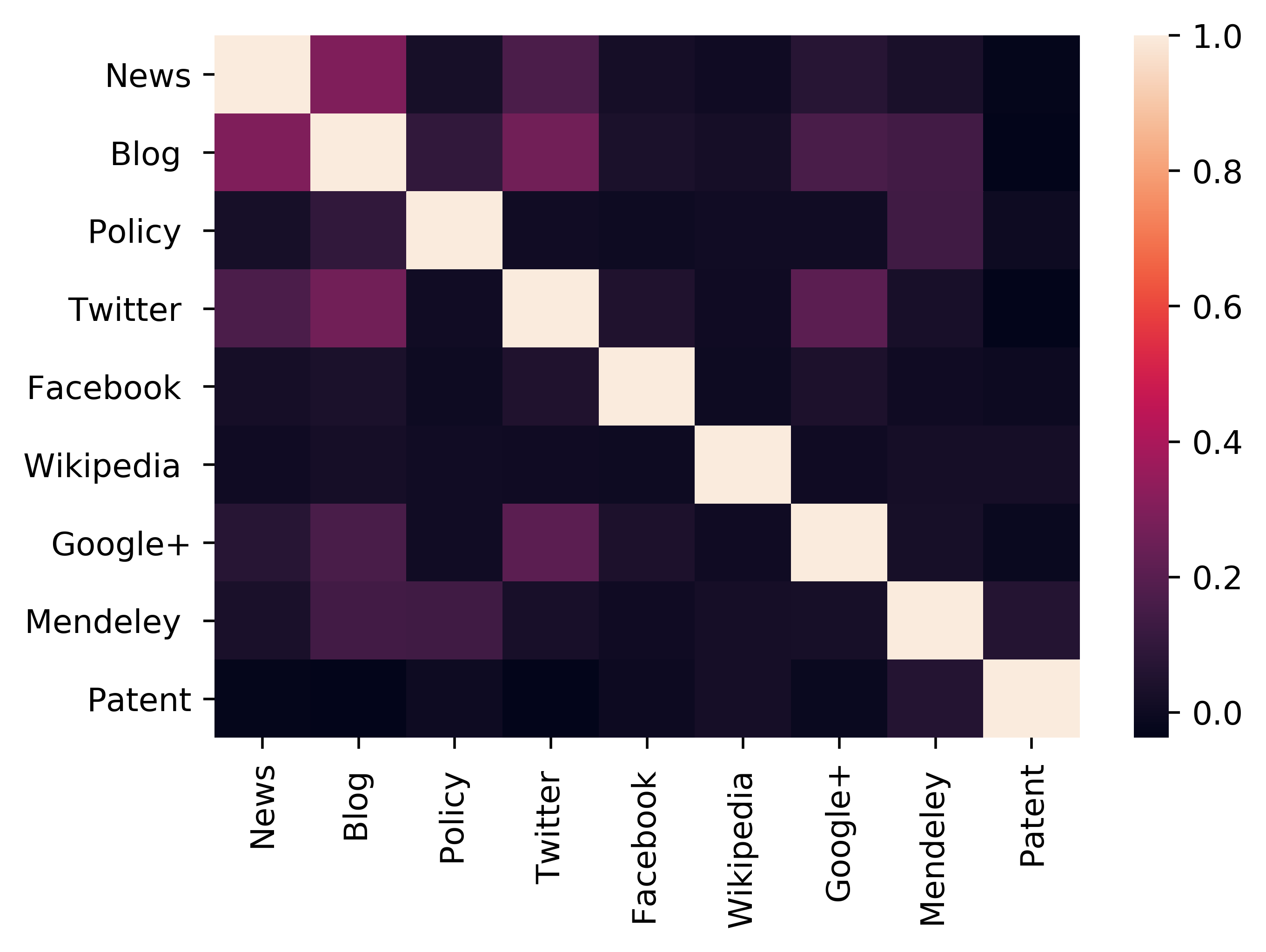}
    \caption{Correlation matrix of features }
    \label{fig:my_label}
\end{figure}

We used social media features since scholarly research is mostly being published or discussed on social media. We had a total of nine features of which one was the target variable and the other eight were predictors used to build classification models. We analyzed the target variable Patent citation and transformed the target variable into binary format such that if a record had a patent citation then the target variable would be 1 and if the patent citation is 0 then the target variable would be 0. The eight predictors were counts of scholarly articles mentions on news outlets, blogs, policy documents, Twitter, Facebook, Wikipedia, Google+ and Mendeley. Those counts were considered before a patent was issued. Fig \ref{fig:my_label} shows the correlation between the features, we can notice that there is very low correlation between the features. We also included the citation count of research papers in the dataset to analyze the relationship between paper citations and patent citations.  

\section{Methodology and Results}
We built four classification models using the processed dataset which are Logistic Regression (LR), Decision Tree (DT), Naive Bayes(NB), and Random Forest (RF). We evaluated the models based on their F1 score and accuracy as shown in Table \ref{Table1}. The table also shows the values of Precision and Recall for the models built on the dataset. After building classification models on the dataset, we observed that Random Forest performed the best in comparison to other models with an accuracy of 93.9\% and an F1-score of 94.5. 

\begin{table}[h]
\begin{tabular}{|l|l|l|l|l|}
\hline
          & LR     & DT     & NB     & RF     \\ \hline
Accuracy  & 89.7\% & 92.6\% & 90.5\% & 93.9\% \\ \hline
F1-score  & 90.3     & 93.0     & 90.4     & 94.5     \\ \hline
Precision & 90.2     & 92.6     & 90.7     & 94.2     \\ \hline
Recall    & 90.4     & 93.4     & 90.1     & 94.8     \\ \hline
\end{tabular}
\vspace{1em}
\caption{Accuracy and F1 based on several machine learning algorithms}
\label{Table1}
\end{table}
\vspace{-1em}

From the dataset of 784,665 records, around 154,270 records had higher than 100 paper citations and among these records 124,267 records had an economic value or were cited by patents. Analyzing this dataset and the correlation matrix we found that papers which were highly cited by other papers were mostly cited by patents as well.

\section{Conclusion and Future Work}
In this paper, we proposed a classification model which to predict whether a research article would have an economic impact by appearing in a patent using various social media features. We also found that papers which were highly cited by other papers were mostly cited by patents as well. In the future, we plan to use market values of patents to analyze the economic impact of a research article and create a framework which could help measure and predict economic impact of a given research article through patents.
%
\bibliographystyle{ACM-Reference-Format}
\bibliography{sample-base}
\end{document}